\documentclass[showpacs,pra]{revtex4} 
\usepackage{amsmath,amssymb} 
\usepackage{graphicx}

\begin{document}

\title{Effective magnetic fields in degenerate atomic gases induced by light
beams with orbital angular momenta}

\author{G. Juzeli\={u}nas$^1$, P. \"{O}hberg$^2$, J. Ruseckas$^{1,3}$ and A.
Klein$^3$}

\affiliation{$^1$Vilnius University Research Institute of Theoretical Physics
and Astronomy,\\
A. Go\v{s}tauto 12, 01108 Vilnius, Lithuania\\
$^2$Department of Physics, University of Strathclyde,\\
Glasgow G4 0NG, Scotland,\\
$^3$Fachbereich Physik der Technischen Universit\"{a}t Kaiserslautern,\\
D-67663 Kaiserslautern, Germany}

\begin{abstract}
We investigate the influence of two resonant laser beams on the
mechanical properties of degenerate atomic gases. The control and probe beams of
light are considered to have Orbital Angular Momenta (OAM) and act on the
three-level atoms in the Electromagnetically Induced Transparency (EIT)
configuration. The theory is based on the explicit analysis of the quantum
dynamics of cold atoms coupled with two laser beams. Using the adiabatic
approximation, we obtain an effective equation of motion for the atoms driven to
the dark state. The equation contains a vector potential type interaction as
well as an effective trapping potential. The effective magnetic field is shown
to be oriented along the propagation direction of the control and probe beams
containing OAM. Its spatial profile can be controlled by choosing proper laser
beams. We demonstrate how to generate a constant effective magnetic field, as
well as a field exhibiting a radial distance dependence. The resulting effective
magnetic field can be concentrated within a region where the effective trapping
potential holds the atoms. The estimated magnetic length can be considerably
smaller than the size of the atomic cloud.
\end{abstract}


\maketitle

\section{Introduction}

During the last decade a remarkable progress has been experienced in trapping
and cooling atoms. In this respect the creation of atomic Bose-Einstein
Condensates (BECs) \cite{ketterle95,hulet95,dalfovo99,bec_stri} and degenerate
Fermi gases \cite{demarco99,salomon01,ketterle03} has been the prime
achievement. The atomic BECs and degenerate Fermi gases are systems where an
atomic physicist often meets physical phenomena encountered in condensed matter
physics. For instance, atoms in optical lattices are often studied using the
Hubbard model \cite{dieter98} familiar from solid state physics.

Atoms forming quantum gases are electrically neutral particles and there is no
vector potential type coupling of the atoms with a magnetic field. Therefore, a
direct analogy between the magnetic properties of degenerate atomic gases and
solid state phenomena is not necessarily straightforward. It is possible to
produce an effective magnetic field in a cloud of electrically neutral atoms by
rotating the system such that the vector potential will appear in the rotating
frame of reference \cite{bretin04,schweikhard04,baranov04}. This would
correspond to a situation where the atoms feel a uniform magnetic field. Yet
stirring an ultracold cloud of atoms in a controlled manner is a rather
demanding task.

There have also been suggestions to take advantage of a discrete periodic
structure of an optical lattice to introduce assymetric atomic transitions
between the lattice sites \cite{jaksch03,mueller04,sorensen04}. Using this
approach one can induce a nonvanishing phase for the atoms moving along a closed
path on the lattice, i.e. one can simulate a magnetic flux
\cite{jaksch03,mueller04,sorensen04}. However such a way of creating the
effective magnetic field is inapplicable to an atomic gas that does not
constitute a lattice.

A significant experimental advantage would be gained if a more direct way could
be used to induce an effective magnetic field. In a previous letter
\cite{prl04}, we have shown how this can be done using two light beams in an
Electromagnetically Induced Transparency (EIT) configuration. Here we present a
more complete account of the phenomenon. We demonstrate that if at least one of
these beams contains an Orbital Angular Momentum (OAM), an effective magnetic
field appears, which acts on the electrically neutral atoms. In other words, the
coupling between the light and the atoms will provide an effective vector
potential in the atomic equations of motion. Compared to the rotating atomic
gas, where only a constant effective magnetic field is created
\cite{bretin04,schweikhard04,baranov04}, using optical means will be
advantageous since the effective magnetic field can now be shaped by choosing
proper control and probe beams. Note that the appearance of our effective vector
potential is a manifestation of the Berry connection which is encountered in
many different areas of physics \cite{jack03,Sun90,Dum96}.

The outline of the paper is as follows. In Sec.~\ref{sec:form} we define a
system of three level atoms in the $\Lambda$-configuration and present the
equations of motion for the atoms interacting with the control and probe beams
of light. In doing this we allow the two beams to have orbital angular momenta
along the propagation axis $ z$. In Sec.~\ref{sec:dark} we derive equations of
motion for the center of mass of atoms driven to the dark state. The equations
of motion contain the terms due to effective vector and trapping potentials
describing an effective magnetic field. In contrast to our previous letter
\cite{prl04}, the emerging effective potentials are now fully Hermitian. Yet,
the two formulations are shown to give the same effective magnetic field and
hence are equivalent. In Secs.~\ref{sec:oam}--\ref{sec:spec} we analyze the
effective magnetic field and effective trapping potential in the case where at
least one of the light beams contains an orbital angular momentum. We show that
the spatial profile of the effective magnetic field can be controlled by
applying proper control and probe beams. The concluding Sec.~\ref{sec:concl}
summarizes the findings. Finally, Appendix \ref{sec:app} contains technical
details of some of the derivations.

\section{\label{sec:form}Formulation}

\subsection{The system}

\begin{figure}
\begin{center}                                                                  
\includegraphics[width=8.5cm]{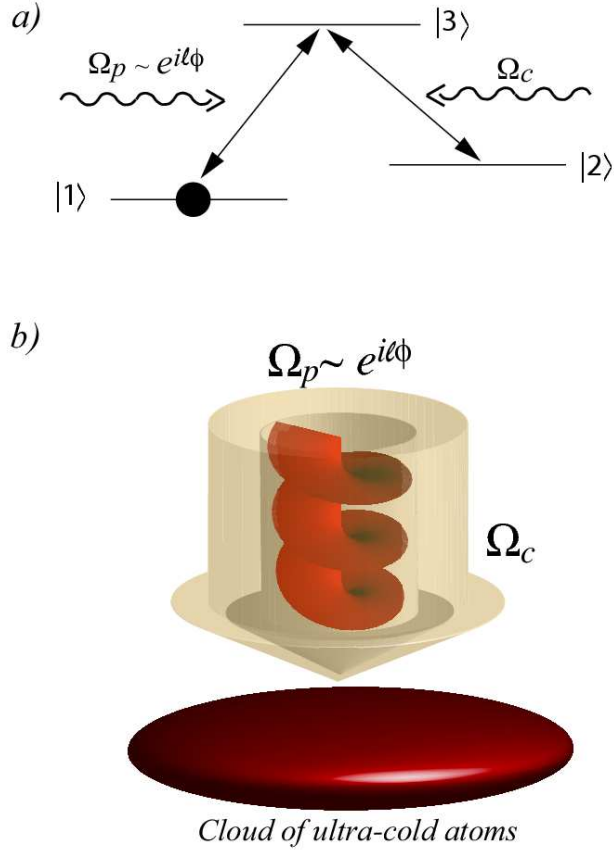}
\end{center}
\caption{(Color online) a) The level scheme for the $\Lambda$
type atoms interacting with the resonant probe beam $\Omega_p$
and control beam $\Omega_c$. b) Schematic representation of the
experimental setup with the two light beams incident on the cloud of atoms. The
probe field is of the form $\Omega_p\sim e^{i\ell\phi}$, where
each probe photon carry an orbital angular momentum $\hbar\ell$
along the propagation axis $ z$.}
\label{BildSchema}
\end{figure}

Let us consider a system of atoms characterized by two hyper-fine ground levels
$ 1$ and $ 2$, as well as an electronic excited level $ 3$. The atoms interact
with two resonant laser beams in the EIT configuration (see
Fig.~\ref{BildSchema}b). The first beam (to be referred to as the control beam)
drives the transition $ |2\rangle\rightarrow |3\rangle$, whereas the second beam
(the probe beam) is coupled with the transition $ |1\rangle\rightarrow
|3\rangle$, see Fig.~\ref{BildSchema}a. The control laser has a frequency
$\omega_c$, a wave-vector $\mathbf{k}_c$, and a Rabi frequency $\Omega_c$. The
probe field, on the other hand, is characterized by a central frequency
$\omega_p=ck_p$, a wave-vector $\mathbf{k}_p$, and a Rabi frequency $\Omega_p$.
Of special interest is the case where the probe and control beams can carry OAM
along the propagation axis $ z$. In that case, the spatial distributions of the
beams are \cite{allen99,oam}
\begin{equation}
\label{omega-p}
\Omega_p=\Omega_p^{(0)}e^{i(k_pz+l_p\phi)}
\end{equation}
 and 
\begin{equation}
\label{omega-c}
\Omega_c=\Omega_c^{(0)}e^{i(k_cz+l_c\phi)}\, ,
\end{equation}
 where $\Omega_p^{(0)}$ and $\Omega_c^{(0)}$ are slowly varying amplitudes for
the probe and control fields, $\hbar\ell_p$ and $\hbar\ell_c$ are the
corresponding orbital angular momenta per photon along the propagation axis $
z$, and $\phi$ is the azimuthal angle.

In first quantization, the quantum mechanical state of the atoms is described in
terms of the three-component wave function $\Psi_j(\mathbf{r},t)$ representing
the probability amplitude to find an atom in the $ j$-th electronic state and
positioned at $\mathbf{r}$, with $ j=1,2,3$ . In second quantization, the
one-particle wave function $\Psi_j(\mathbf{r},t)$ is replaced by the operator
$\Psi_j(\mathbf{r},t)$ for annihilation of an atom positioned at $\mathbf{r}$
and characterized by an internal state $ j$. A set of such operators
$\Psi_j(\mathbf{r},t)$ obeys the Bose-Einstein or Fermi-Dirac commutation
relationships depending on the type of atoms involved. In what follows,
$\Psi_j(\mathbf{r},t)$ can be understood either as the three-component atomic
wave function or as the annihilation field operator. In both cases the spatial
and temporal variables will be kept implicit,
$\Psi_j(\mathbf{r},t)\equiv\Psi_j$.

\subsection{Initial equations of motion}

Introducing the slowly-varying atomic field-operators
$\Phi_1=\Psi_1e^{i\omega_1t}$, $\Phi_3=\Psi_3e^{i(\omega_1+\omega_p)t}$ and
$\Phi_2=\Psi_2e^{i(\omega_1+\omega_p-\omega_c)t}$ and adopting the rotating wave
approximation, the equations of motion read 
\begin{eqnarray}
i\hbar\dot{\Phi}_1 & = & -\frac{\hbar^2}{2m}\nabla^2\Phi_1+V_1(\mathbf{r})\Phi_1
+\hbar\Omega_p^*\Phi_3\, ,
\label{eq-at-g2}\\
i\hbar\dot{\Phi}_3 & = & -\frac{\hbar^2}{2m}\nabla^2\Phi_3+[\epsilon_{31}
+V_3(\mathbf{r})]\Phi_3+\hbar\Omega_c\Phi_2+\hbar\Omega_p\Phi_1\, ,
\label{eq-at-e2}\\
i\hbar\dot{\Phi}_2 & = & -\frac{\hbar^2}{2m}\nabla^2\Phi_2+[\epsilon_{21}
+V_2(\mathbf{r})]\Phi_2+\hbar\Omega_c^*\Phi_3\, ,
\label{eq-at-q2}
\end{eqnarray}
 where $ m$ is the atomic mass, $ V_j(\mathbf{r})$ is the trapping potential for
an atom in the electronic state $ j$,
$\epsilon_{21}=\hbar(\omega_2-\omega_1+\omega_c-\omega_p)$ and
$\epsilon_{31}=\hbar(\omega_3-\omega_1-\omega_p)$ are, respectively, the
energies of the detuning from the two- and single-photon resonances with
$\hbar\omega_j$ being the electronic energy of the atomic level $ j$.

The equations of motion (\ref{eq-at-g2})-(\ref{eq-at-q2}) do not accommodate
collisions between the ground-state atoms. This is a legitimate approximation
for a degenerate Fermi gas in which s-wave scattering is forbidden and only weak
p-wave scattering is present \cite{butts97,demarco99,mewes00,juzeliunas01}. On
the other hand, if the atoms in the hyperfine ground states $ 1$ and $ 2$ form a
BEC, collisions will be present between these atoms. The collisional interaction
can however be accommodated if Eqs.~(\ref{eq-at-g2}) and (\ref{eq-at-q2}) are
replaced by the mean-field (Gross-Pitaevskii) equations for the condensate
wave-functions,
\begin{eqnarray}
i\hbar\dot{\Phi}_1 & = &\left(-\frac{\hbar^2}{2m}\nabla^2V_1(\mathbf{r})+g_{
11}|\Phi_1|^2+g_{12}|\Phi_2|^2\right)\Phi_1+\hbar\Omega_p^*\Phi_3\, ,
\label{eq-at-g2-interact}\\
i\hbar\dot{\Phi}_2 & = &\left(-\frac{\hbar^2}{2m}\nabla^2\epsilon_{21}
+V_2(\mathbf{r})+g_{12}|\Phi_1|^2+g_{22}|\Phi_2|^2\right)\Phi_2
+\hbar\Omega_c^*\Phi_3\, ,
\label{eq-at-q2-interact}
\end{eqnarray}
 where $ g_{jl}=4\pi\hbar^2a_{jl}/m$, with $ a_{jl}$ the s-wave scattering
length of the atoms in the electronic states $ j$ and $ l$, respectively. In
particular $ a_{jj}$ is the length of the s-wave scattering between a pair of
atoms in the same electronic state ($ j=1,2$), whereas $ a_{12}=a_{21}$
corresponds to collisions between atoms in different electronic states. Since
the occupation of the excited atomic level $ 3$ is small, the atom-atom
scattering is of little importance for these atoms and equation (\ref{eq-at-e2})
for $\Phi_3$ can therefore be left unaltered.

\section{\label{sec:dark}Dark state representation}

\subsection{Transformed equations of motion}

It is convenient to introduce the annihilation field operators for the atoms in
the dark and bright states,
\begin{eqnarray}
\Phi_D & = &\frac{1}{\sqrt{1+|\zeta |^2}}(\Phi_1-\zeta^*\Phi_2)\, ,
\label{Psi-D}\\
\Phi_B & = &\frac{1}{\sqrt{1+|\zeta |^2}}(\zeta\Phi_1+\Phi_2)\, ,
\label{Psi-B}
\end{eqnarray}
 where
\begin{equation}
\zeta =\frac{\Omega_p}{\Omega_c}
\end{equation}
is the ratio of the amplitudes of the control and probe fields.

We shall be especially interested in a situation where the atoms are driven to
their dark states, described by the creation field operator 
$\Phi_D^{\dag} (\mathbf{r},0) $ acting on the atomic vacuum $|\mathrm{vac}\rangle$. 
If an atom is in the dark state
$|D\rangle\sim |1\rangle -\zeta |2\rangle$, the
resonant control and probe beams induce the absorption paths $
|2\rangle\rightarrow |3\rangle$ and $ |1\rangle\rightarrow |3\rangle$ which
interfere destructively, resulting in the Electromagnetically Induced
Transparency \cite{Arimondo96,Harris97,eit,lukin03}. In fact, as one can see
from Eq.~(\ref{eq-at-e2}), the transitions to the upper atomic level $ 3$ are
then suppressed, so the atomic level $ 3$ is weakly populated. This justifies
neglection of losses due to spontaneous emissions by the excited atoms in the
equation (\ref{eq-at-e2}) for $\Phi_3$.

A transformed set of operators $\Phi_D$, $\Phi_B$ and $\Phi_3$ obeys the
following equations of motion (see Appendix A): 
\begin{eqnarray}
i\hbar\dot{\Phi}_D & = &\frac{1}{2m}\left(i\hbar\nabla +\mathbf{A}_{\mathrm{
eff}}^{(D)}\right)^2\Phi_D+V_{\mathrm{eff}}^{(D)}(\mathbf{r})\Phi_D+F_{
DB}(\mathbf{r})\Phi_B\, ,
\label{eq:Psi-D}\\
i\hbar\dot{\Phi}_B & = &\frac{1}{2m}\left(i\hbar\nabla +\mathbf{A}_{\mathrm{
eff}}^{(B)}\right)^2\Phi_B+V_{\mathrm{eff}}^{(B)}(\mathbf{r})\Phi_B
+\hbar\Omega\Phi_3+F_{BD}(\mathbf{r})\Phi_D\, ,
\label{eq:Psi-B}\\
i\hbar\dot{\Phi}_3 & = & -\frac{\hbar^2}{2m}\nabla^2\Phi_3+[\epsilon_{31}
+V_3(\mathbf{r})]\Phi_3+\hbar\Omega\Phi_B\, ,
\label{eq:Psi3}
\end{eqnarray}
 where
\begin{equation}
\label{Omega-R}
\Omega(\mathbf{r})=\sqrt{|\Omega_p|^2+|\Omega_c|^2}
\end{equation}
 is the total Rabi frequency, 
\begin{equation}
\label{A-eff-D}
\mathbf{A}_{\mathrm{eff}}^{(D)}=-\mathbf{A}_{\mathrm{eff}}^{(B)}=\frac{i\hbar}{
2}\frac{\zeta^*\nabla\zeta -\zeta\nabla\zeta^*}{1+|\zeta |^2}
\end{equation}
 is the \textit{effective vector potential} and 
\begin{eqnarray}
V_{\mathrm{eff}}^{(D)}(\mathbf{r}) & = &\frac{1}{1+|\zeta |^2}\left(V_1(\mathbf{
r})+|\zeta |^2(V_2(\mathbf{r})+\epsilon_{21})\right)-\frac{i\hbar}{2}\frac{
\zeta^*\dot{\zeta}-\zeta\dot{\zeta}^*}{1+|\zeta |^2}\nonumber\\
 &  & +\frac{\hbar^2}{2m}\frac{\nabla\zeta^*\nabla\zeta}{\left(1+|\zeta
|^2\right)^2}\, ,
\label{V-eff-D}\\
V_{\mathrm{eff}}^{(B)}(\mathbf{r}) & = &\frac{1}{1+|\zeta |^2}\left(|\zeta
 |^2V_1(\mathbf{r})+V_2(\mathbf{r})+\epsilon_{21}\right)-\frac{i\hbar}{2}\frac{
\dot{\zeta}^*\zeta -\zeta^*\dot{\zeta}}{1+|\zeta |^2}\nonumber\\
 &  & +\frac{\hbar^2}{2m}\frac{\nabla\zeta^*\nabla\zeta}{\left(1+|\zeta
 |^2\right)^2}
\label{V-eff-B}
\end{eqnarray}
are the \textit{effective trapping potentials} for the atoms in the dark and
bright states, respectively. The operators $ F_{DB}$ and $ F_{BD}$ describing
transitions between the dark and bright states in Eqs.~(\ref{eq:Psi-D}) and
(\ref{eq:Psi-B}), are explicitly defined in Appendix A. Note that the effective
vector and trapping potentials $\mathbf{A}_{\mathrm{eff}}^{(D)}$ and $
V_{\mathrm{eff}}^{(D)}(\mathbf{r})$ are Hermitian.

The effective magnetic field, corresponding to the effective vector potential
$\mathbf{A}_{\mathrm{eff}}^{(D)}$ is 
\begin{equation}
\label{B-eff}
\mathbf{B}_{\mathrm{eff}}=\nabla\times\mathbf{A}_{\mathrm{eff}}^{
(D)}=i\hbar\frac{1}{\left(1+|\zeta |^2\right)^2}\nabla\zeta^*\times\nabla\zeta\,
 .
\end{equation}

\subsection{Equation of motion under adiabatic approximation}

In what follows we shall restrict ourselves to the \textit{adiabatic} case in
which transitions between the dark and bright states are not important. In such
a situation the term $ F_{DB}$ can be neglected in Eq.~(\ref{eq:Psi-D}), so it
is sufficient to consider a single equation describing the translational motion
of the atoms in the dark state: 
\begin{equation}
\label{eq:Psi-D-only}
i\hbar\dot{\Phi}_D=\frac{1}{2m}\left(i\hbar\nabla +\mathbf{A}_{\mathrm{eff}}^{
(D)}\right)^2\Phi_D+V_{\mathrm{eff}}^{(D)}(\mathbf{r})\Phi_D\, .
\end{equation}
 Assuming that the control and probe fields are tuned to the one- and two-photon
resonances ($\epsilon_{31},\epsilon_{21}\ll\hbar\Omega$), the adiabatic approach
holds if the matrix elements of the operators $ F_{DB}$ and $ F_{BD}$ are much
smaller than the total Rabi frequency $\Omega$. This leads to the following
requirement for the velocity-dependent term in $ F_{DB}$ 
\begin{equation}
\label{condition-adiabatic}
F\ll\Omega\, .
\end{equation}
Here the velocity-dependent term 
\begin{equation}
\label{F}
F=\frac{1}{1+|\zeta |^2}\left|\nabla\zeta\cdot\mathbf{v}\right|
\end{equation}
 reflects the two-photon Doppler detuning. Note that the estimation
(\ref{condition-adiabatic}) does not accommodate effects due to the decay of the
excited atoms. The dissipation effects can be included replacing the energy of
the one-photon detuning $\epsilon_{31}$ by $\epsilon_{31}-i\hbar\gamma_3$, where
$\gamma_3$ is the excited-state decay rate. In such a situation, the dark state
can be shown to acquire a finite lifetime 
\begin{equation}
\label{Gamma-D}
\tau_D\sim\gamma_3^{-1}\Omega^2/F^2
\end{equation}
 which should be large compared to other characteristic times of the system. The
adiabatic conditions will be further analyzed in Subsection \ref{sec:adiab2}.

If the atoms in the hyperfine ground states $ 1$ and $ 2$ form a BEC, the atomic
dynamics in these states is governed by the mean field equations
(\ref{eq-at-g2-interact})-(\ref{eq-at-q2-interact}). In such a situation, the
equation of motion for the dark state atoms modifies as 
\begin{equation}
\label{eq:Psi-D-interact}
i\hbar\dot{\Phi}_D=\frac{1}{2m}\left(i\hbar\nabla +\mathbf{A}_{\mathrm{eff}}^{
(D)}\right)^2\Phi_D+V_{\mathrm{eff}}^{(D)}(\mathbf{r})\Phi_D
+g_D|\Phi_D|^2\Phi_D\, ,
\end{equation}
 where 
\begin{equation}
\label{g-D}
g_D=\frac{1}{\left(1+|\zeta |^2\right)^2}(g_{11}+2g_{12}|\zeta |^2+|\zeta |^4g_{
22})
\end{equation}
 describes the interaction between the atoms in the dark state.

\subsection{Relation to previous work}

In our previous letter \cite{prl04} an effective equation of motion has been
derived for the atoms in the hyperfine ground level $ 1$. In doing this, the
atoms were assumed to be driven to their dark states by imposing the constraint
$\Phi_2(\mathbf{r},t)=-\zeta\Phi_1(\mathbf{r},t)$, which is equivalent to the
requirement $\Phi_B(\mathbf{r},t)=0$. The resulting effective equation of motion
for $\Phi_1(\mathbf{r},t)$ reads \cite{prl04}: 
\begin{equation}
\label{eq:psi-1}
i\hbar\dot{\Phi}_1=\frac{1}{2m}\left[i\hbar\nabla +\mathbf{A}_{\mathrm{
eff}}\right]^2\Phi_1+V_{\mathrm{eff}}(\mathbf{r})\Phi_1\, ,
\end{equation}
 where the effective vector and trapping potentials are generally non-Hermitian.
For instance, the effective vector potential featured in Eq.~(\ref{eq:psi-1}) is
given by \cite{prl04}: 
\begin{equation}
\label{A-eff-rez}
\mathbf{A}_{\mathrm{eff}}=\frac{i\hbar\zeta^*\nabla\zeta}{1+|\zeta
 |^2}\equiv\mathbf{A}_{\mathrm{eff}}^{(D)}-i\hbar\nabla\ln\left(1+|\zeta
 |^2\right)^{-1/2}\, .
\end{equation}
 Non-Hermitian potentials appear because the atoms in the electronic state $ 1$
constitute an open sub-system. In fact, the probe and control beams transfer
reversibly atomic population from level $ 1$ to level $ 2$ by means of the
two-photon Raman transition.

Using the constraint $\Phi_2=-\zeta\Phi_1$, one can express the dark-state
operator $\Phi_D$ given by Eq.~(\ref{Psi-D}) in terms of $\Phi_1$ as: 
\begin{equation}
\label{Psi-D-Psi-1}
\Phi_D=\Phi_1\left(1+|\zeta |^2\right)^{1/2}\equiv\Phi_1\exp\left[\ln\left(1
+|\zeta |^2\right)^{1/2}\right]\, .
\end{equation}
 Equation (\ref{Psi-D-Psi-1}) represents a pseudo-gauge (non-unitary)
transformation relating the effective equation of motion (\ref{eq:psi-1}) for
$\Phi_1$ to the corresponding equation for the dark-state operator $\Phi_D$. The
transformation (\ref{Psi-D-Psi-1}) is not unitary as long as the intensity of
the probe field is non-zero ($ |\zeta |\neq 0$). The transition from the unitary
equation of motion for $\Phi_D$ to the non-unitary one for $\Phi_1$ is
accompanied by of the non-Hermitian vector and trapping potentials
$\mathbf{A}_{\mathrm{eff}}$ and $ V_{\mathrm{eff}}$. The Hermitian potential
$\mathbf{A}_{\mathrm{eff}}^{(D)}$ differs from its non-Hermitian counterpart
$\mathbf{A}_{\mathrm{eff}}$ by a gradient of the imaginary function $
i\hbar\ln(1+|\zeta |^2)^{-1/2}$, as one can see from Eq.~(\ref{eq:psi-1}). In a
similar manner, the Hermitian trapping potential $
V_{\mathrm{eff}}^{(D)}(\mathbf{r})$ can be shown to differ from the
non-Hermitian potential $ V_{\mathrm{eff}}(\mathbf{r})$ by the time-derivative
of the imaginary function $ -i\hbar\ln(1+|\zeta |^2)^{-1/2}$. In this way the
two formulations are equivalent. Since $\mathbf{A}_{\mathrm{eff}}^{(D)}$ differs
from $\mathbf{A}_{\mathrm{eff}}$ by a gradient, the effective magnetic field
{[}Eq.~(\ref{B-eff}){]} acting on the dark-state atoms, is the same in both
formulations.

\section{\label{sec:oam}Effective potentials due to light beams with OAM}

\subsection{Representation in terms of the amplitude and phase}

Separating the ratio $\zeta$ into an amplitude and phase, 
\begin{equation}
\label{ampl-phase}
\zeta =\Omega_p/\Omega_c=|\zeta |e^{iS}\, ,
\end{equation}
 the effective vector and trapping potentials given by Eqs.~(\ref{A-eff-D}) and
(\ref{V-eff-D}) can be rewritten as 
\begin{equation}
\label{A-eff-D-ampl-phase}
\mathbf{A}_{\mathrm{eff}}^{(D)}=-\hbar\frac{|\zeta |^2}{1+|\zeta |^2}\nabla S
\end{equation}
 and 
\begin{equation}
\label{V-eff-D-ampl-phase}
V_{\mathrm{eff}}^{(D)}(\mathbf{r})=V_{\mathrm{ext}}(\mathbf{r})+\frac{\hbar^2}{
2m}\frac{|\zeta |^2(\nabla S)^2+(\nabla |\zeta |)^2}{\left(1+|\zeta
 |^2\right)^2}+\frac{|\zeta |^2(\hbar\dot{S}+\epsilon_{21})}{1+|\zeta |^2}\, ,
\end{equation}
where
\begin{equation}
\label{vext}
V_{\mathrm{ext}}(\mathbf{r})=\frac{V_1(\mathbf{r})+|\zeta |^2V_2(\mathbf{r})}{1
+|\zeta |^2}
\end{equation}
 is the \textit{external trapping potential} for the atoms in the dark state.
The effective magnetic field then takes the form 
\begin{equation}
\label{B-eff-D-ampl-phase}
\mathbf{B}_{\mathrm{eff}}=\hbar\frac{(\nabla S)\times\nabla |\zeta |^2}{\left(1
+|\zeta |^2\right)^2}\, ,
\end{equation}
i.e. the strength of the effective magnetic field is determined by the cross
product of the gradients of the amplitude and phase $ (\nabla S)\times\nabla
|\zeta |^2$.

\subsection{Control and probe beams with OAM}

If the co-propagating probe and control fields carry OAM, their amplitudes
$\Omega_p$ and $\Omega_c$ are given by Eqs.~(\ref{omega-p})--(\ref{omega-c}).
The phase of the ratio $\zeta =\Omega_p/\Omega_c$ then reads 
\begin{equation}
\label{S}
S=l\phi\, ,
\end{equation}
 where $ l=l_p-l_c$. Note that although both the control and probe fields are
generally allowed to have non-zero OAM by Eqs.~(\ref{omega-p})--(\ref{omega-c}),
it is desirable that the OAM is zero for one of these beams. In fact, if both $
l_p$ and $ l_c$ were non-zero, the amplitudes $\Omega_p$ and $\Omega_c$ should
simultaneously go to zero along the $ z$-axis. In such a situation, the total
Rabi frequency $\Omega =(\Omega_p^2+\Omega_c^2)^{1/2}$ would also vanish,
leading to the violation of the adiabatic condition (\ref{condition-adiabatic})
along the $ z$-axis.

Substituting Eq.~(\ref{S}) into Eqs.~(\ref{A-eff-D-ampl-phase}) and
(\ref{B-eff-D-ampl-phase}), the effective vector potential and magnetic field
take the form 
\begin{eqnarray}
\mathbf{A}_{\mathrm{eff}}^{(D)} & = & -\frac{\hbar l}{\rho}\frac{|\zeta |^2}{1
+|\zeta |^2}\hat{\mathbf{e}}_{\phi}\, ,
\label{A-eff-D-OAM}\\
\mathbf{B}_{\mathrm{eff}} & = &\frac{\hbar l}{\rho}\frac{1}{\left(1
+\left|\zeta\right|^2\right)^2}\hat{\mathbf{e}}_{\phi}\times\nabla
|\zeta |^2\, ,
\label{B-eff-D-ampl-phase-1}
\end{eqnarray}
where $\rho$ is the cylindrical radius and $\hat{\mathbf{e}}_{\phi}$
is the unit vector along the azimuthal angle $\phi$. In a similar manner, with
the electronic two photon detuning put to zero ($\epsilon_{21}=0$),
Eq.~(\ref{V-eff-D-ampl-phase}) reduces to 
\begin{equation}
\label{V-eff-D-OAM}
V_{\mathrm{eff}}^{(D)}(\mathbf{r})=V_{\mathrm{ext}}(\mathbf{r})+\frac{\hbar^2}{
2m}\frac{l^2|\zeta |^2/\rho^2+(\nabla |\zeta |)^2}{\left(1+|\zeta
 |^2\right)^2}\, .
\end{equation}

In what follows we shall assume that the intensity ratio $ |\zeta |^2$ depends
on the \textit{cylindrical radius} $\rho$ only. In that case the effective
magnetic field is directed along the z-axis
\begin{equation}
\label{B-eff-OAM}
\mathbf{B}_{\mathrm{eff}}=-\hat{\mathbf{e}}_z\frac{\hbar l}{\rho}\frac{1}{
\left(1+\left|\zeta\right|^2\right)^2}\frac{\partial}{\partial\rho}|\zeta |^2\,
 .
\end{equation}
It is evident that the effective magnetic field is non-zero only if the ratio
$\zeta =\Omega_p/\Omega_c$ contains a non-zero phase ($ l=l_p-l_c\neq 0$) and
the amplitude $ |\zeta |$ has a radial dependence ($\partial |\zeta
|/\partial\rho\neq 0$).

\subsection{\label{sec:adiab2}Adiabatic condition}

For light beams with OAM the \textit{adiabatic condition} given by
Eq.~(\ref{condition-adiabatic}) can be rewritten as 
\begin{equation}
\label{condition-adiabatic-1}
\frac{1}{1+|\zeta |^2}\sqrt{\left(v_{\rho}\frac{\partial}{\partial\rho}|\zeta
 |\right)^2+\left(|\zeta |v_{\phi}\frac{l}{\rho}\right)^2}\ll\Omega\, .
\end{equation}
 The above condition imposes requirements on the radial and azimuthal atomic
velocities $ v_{\rho}$ and $ v_{\phi}=\rho\omega_{\phi}$, where $\omega_{\phi}$
is an angular frequency of the atomic motion. Note that condition
(\ref{condition-adiabatic-1}) has no singularity due to the $\rho^{-1}$ term,
since for the light beams with OAM the ratio $ |\zeta
|=\left|\Omega_p/\Omega_c\right|$ typically goes as $\rho^l$ close to the origin
\cite{oam}.

The condition (\ref{condition-adiabatic-1}) implies that the inverse Rabi
frequency $\Omega^{-1}$ should be smaller than the time an atom travels a
characteristic length over which the amplitude or the phase of the ratio $\zeta
=\Omega_p/\Omega_c$ changes considerably. The latter length exceeds the optical
wavelength, and the Rabi frequency can be of the order of $ 10^7$ to $
10^8\,\mathrm{s}^{-1}$ \cite{hau99}. Consequently, the adiabatic condition
(\ref{condition-adiabatic-1}) should still hold for atomic velocities of the
order of tens of meters per second, i.e., up to extremely large velocities in
the context of ultra-cold atomic gases. The allowed atomic velocities become
lower if the spontaneous decay of the excited atoms is taken into account.
According to Eq.~(\ref{Gamma-D}), the atomic dark state accquires then a finite
lifetime $\tau_D$ which is determined by $\gamma_3^{-1}$ times the ratio
$\Omega^2/F^2$. The atomic decay rate $\gamma_3$ is typically of the order $
10^7\,\mathrm{s}^{-1}$. Therefore, in order to achieve long-lived dark states
the atomic speed should not be too large. For instance, if the atomic velocities
are of the order of a centimeter per second (a typical speed of sound in a BEC),
the atoms should survive in their dark states up to a few seconds. This is
comparable to the typical lifetime of an atomic BEC.

\section{\label{sec:spec}Specific cases}

Suppose the probe beam has an OAM ($ l_p\neq 0$) and the control beam does not
($ l_c=0$). In this case the intensity of the probe beam (and hence the ratio $
|\zeta |^2=|\Omega_p/\Omega_c|^2)$ goes to zero as $\rho\rightarrow 0$. If the
intensity of the control field changes slowly within an atomic cloud, the
$\rho$-dependence of the ratio $ |\zeta |$ is determined by the probe beam only.

The effective magnetic flux through a circle of the radius $\rho_0$ is now given
by 
\begin{equation}
\label{flux}
\Phi =\oint\mathbf{A}_{\mathrm{eff}}^{(D)}\mathrm{d}\mathbf{l}=-2\pi\hbar\frac{
l|\zeta_0|^2}{1+|\zeta_0|^2}\, ,
\end{equation}
 where $ 2\pi\hbar$ is the Dirac flux quantum, and $ |\zeta_0|^2$ is the
intensity ratio at the radius $\rho =\rho_0$. The flux $\Phi$ reaches its
maximum of $ 2\pi\hbar l$ if the ratio $ |\zeta_0|^2\gg 1$, i.e. if the
intensity of the probe field exceeds the control field at the selected radius
$\rho_0$. Since the winding number of light beams can currently be as large as
several hundreds, it is possible to induce a substantial flux $\Phi$ in the
atomic cloud. This might enable us to study phenomena related to filled Landau
levels with a large number of atoms in the quantum gases.

\subsection{The case where $ |\zeta |\sim\rho^n$}

Let us consider the case where the probe beam containing an OAM exhibits the
power law behaviour $ |\zeta |=\alpha\rho^n$. Under this condition,
Eqs.~(\ref{V-eff-D-OAM}) and (\ref{B-eff-OAM}) take the form 
\begin{equation}
\label{B-eff-OAM-2}
\mathbf{B}_{\mathrm{eff}}=-2nl\hbar\frac{\alpha^2\rho^{2n-2}}{\left(1
+\alpha^2\rho^{2n}\right)^2}\hat{\mathbf{e}}_z
\end{equation}
 and 
\begin{equation}
\label{V-eff-D-OAM-parax}
V_{\mathrm{eff}}^{(D)}(\mathbf{r})=V_{\mathrm{ext}}(\mathbf{r})+\frac{\hbar^2}{
2m}\frac{\left[l^2+n^2\right]\alpha^2\rho^{2n-2}}{\left(1+\alpha^2\rho^{
2n}\right)^2}\, .
\end{equation}
 If the probe beam is characterised by a winding number $ l_p=l$, the radial
distribution typically goes as $ |\zeta |=\alpha\rho^l$ for small values of
$\rho$ \cite{oam}. Therefore, for $ l>1$, the effective magnetic field goes to
zero at the origin, where $\rho =0$. It is desirable to exclude this area by
introducing a repulsive potential expelling the atoms for small values of
$\rho$. In what follows we shall consider some other types of radial dependence
which are relevant for a larger cylindrical radius $\rho$.

\subsection{The case where $ |\zeta |$ is linear in $\rho$}

If $ |\zeta |=\alpha\rho$, we get 
\begin{equation}
\label{B-eff-OAM-3}
\mathbf{B}_{\mathrm{eff}}=-2l\hbar\hat{\mathbf{e}}_z\frac{\alpha^2}{\left(1
+\alpha^2\rho^2\right)^2}
\end{equation}
 and 
\begin{equation}
\label{V-eff-D-OAM-parax-1}
V_{\mathrm{eff}}^{(D)}(\mathbf{r})=V_{\mathrm{ext}}(\mathbf{r})+\frac{\hbar^2}{
2m}\frac{\left[l^2+1\right]\alpha^2}{\left(1+\alpha^2\rho^2\right)^2}\, .
\end{equation}
 For sufficiently small distances ($ |\zeta |=\alpha\rho\ll 1$),
Eq.~(\ref{B-eff-OAM-3}) describes a constant magnetic field along the $ z$-axis,
in agreement with Eq.~(11) of Ref.~\cite{prl04}. Retaining terms up to quadratic
order in $\rho$, the effective trapping potential,
Eq.~(\ref{V-eff-D-OAM-parax-1}), becomes 
\begin{equation}
\label{V-eff-D-OAM-parax-2}
V_{\mathrm{eff}}^{(D)}(\mathbf{r})\approx V_{\mathrm{ext}}(\mathbf{r})+\frac{
\hbar^2}{2m}\left[l^2+1\right]\alpha^2(1-2\alpha^2\rho^2)\, .
\end{equation}
Assuming
\begin{equation}
\label{V-1}
V_1(\mathbf{r})=\frac{\hbar^2}{m}\left[l^2+1\right]\alpha^4\rho^2
\end{equation}
 and $ V_2(\mathbf{r})=\kappa V_1(\mathbf{r})$, the external trapping potential
$ V_{\mathrm{ext}}(\mathbf{r})$, Eq.~(\ref{vext}), compensates the quadratic
distance dependence in the second term of Eq.~(\ref{V-eff-D-OAM-parax-2}). In
such a situation, the overall effective potential $
V_{\mathrm{eff}}^{(D)}(\mathbf{r})$ is constant up to terms of the fourth order
in $\rho$.

\begin{figure}
\begin{center}                                                                  
\includegraphics[width=8.5cm,angle=0]{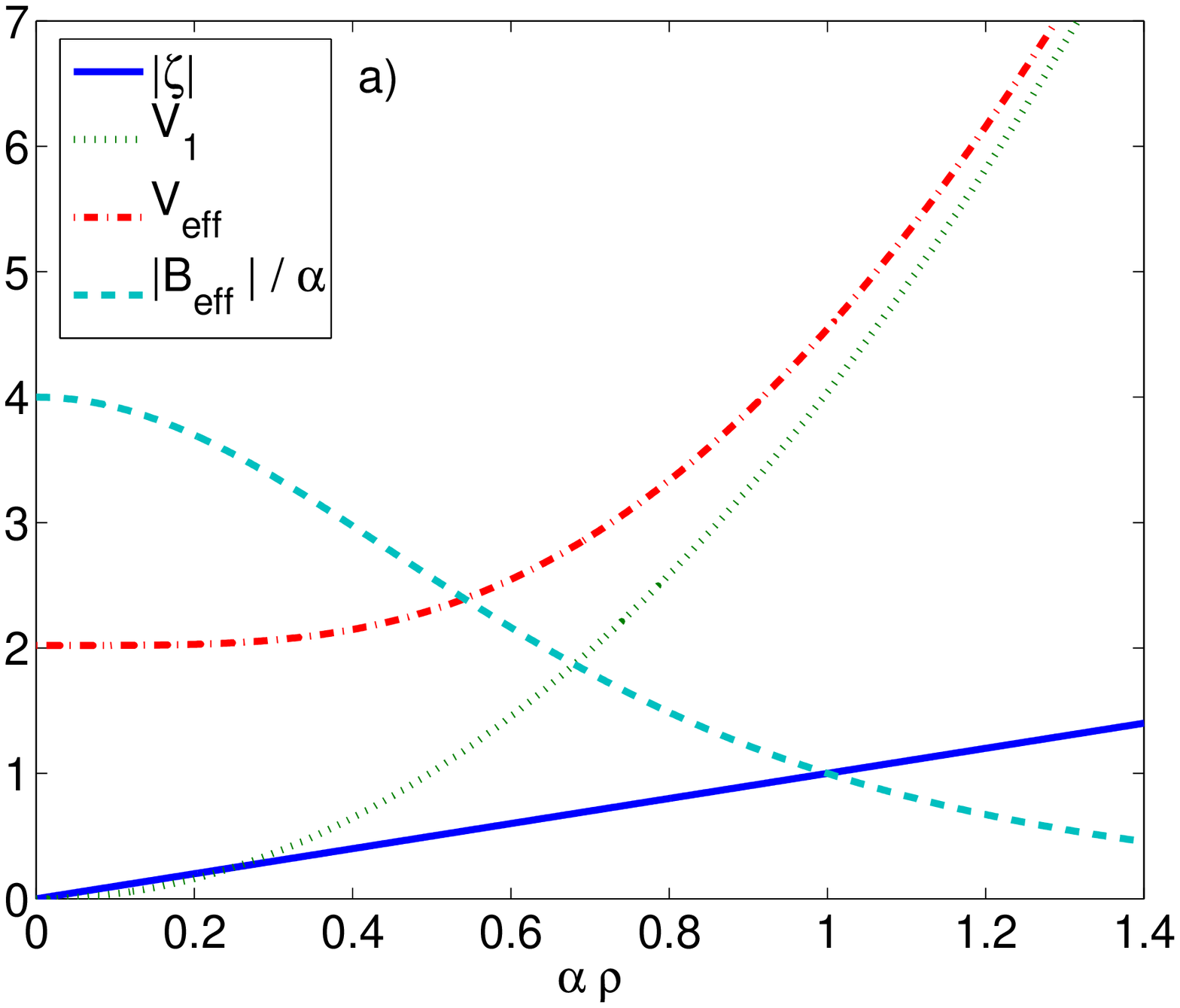}                             
\includegraphics[width=8.5cm,angle=0]{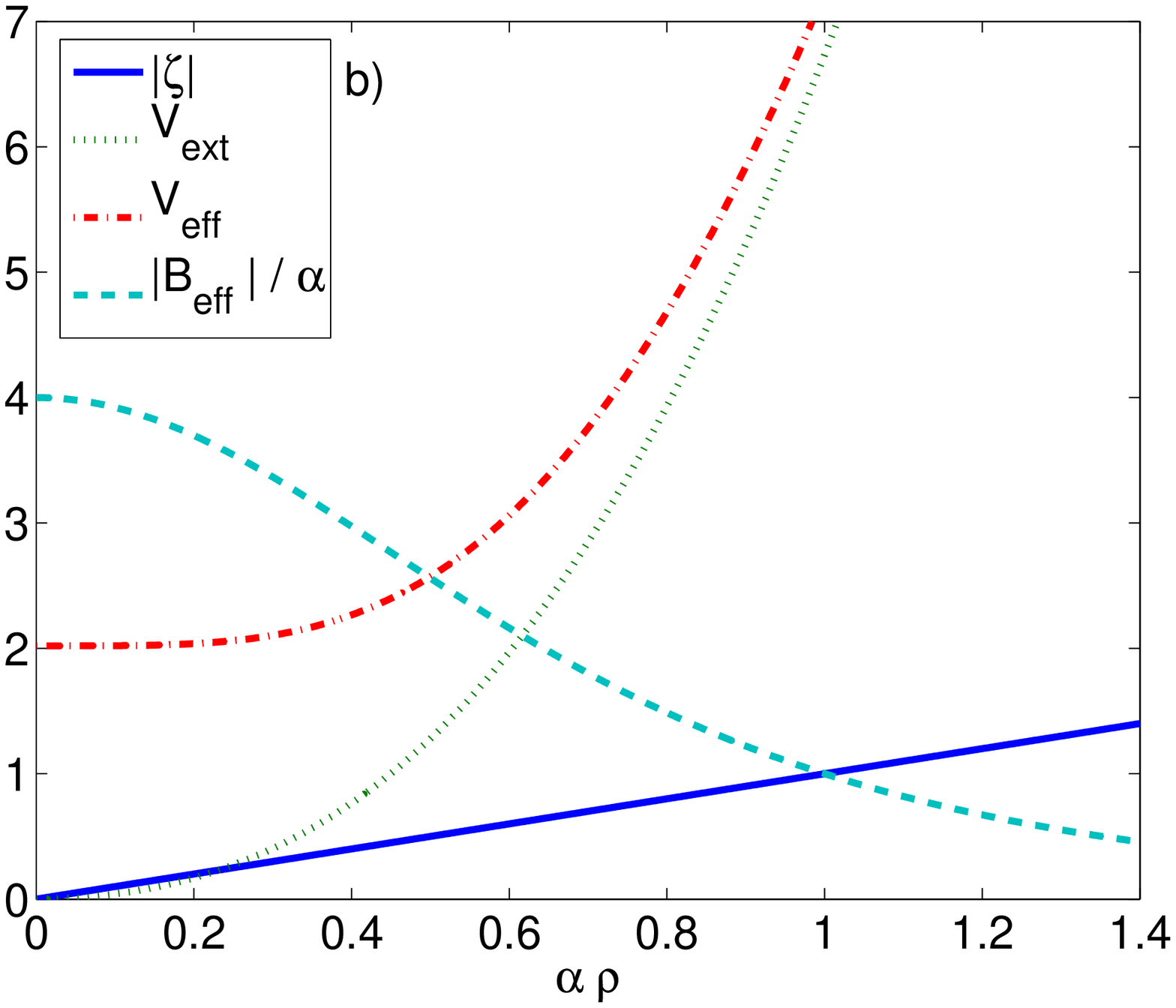}                         
\end{center}
\caption{(Color online) Effective trapping potential $
V_{\mathrm{eff}}$ and effective magnetic field $
B_{\mathrm{eff}}$ for the case where $ |\zeta |$ is
linear in $\rho$ and the constants are $ m=\hbar
=1$, $\alpha =0.2$, $ l=10$. The
trapping potential $ V_1$ is chosen to be given by
Eq.~(\ref{V-1}), so that the quadratic term of the effective trapping potential
vanishes. The trapping potential for the atoms in the second hyperfine ground
state is chosen to be $ V_2(\mathbf{r})=\kappa V_1(\mathbf{r})$
with (a) $\kappa =1$ and (b) $\kappa =7/3$.}
\label{fig:2}
\end{figure}

Figure \ref{fig:2} shows the effective magnetic field and the trapping potential
for the whole range of distances $\rho$ in the case where $
V_2(\mathbf{r})=\kappa V_1(\mathbf{r})$, with $\kappa =1$ (Fig.~\ref{fig:2}a)
and $\kappa =7/3$ (Fig.~\ref{fig:2}b). The external trapping potential is
defined here by Eqs.~(\ref{vext}) and (\ref{V-1}). The overall trapping
potential is seen to be flat for small distances ($\alpha\rho\ll 1$). In this
area the magnetic field is close to its maximum value. For larger distances an
effective trapping barrier is formed preventing the atoms to escape the area
where the magnetic field is contained, as seen in Fig.~\ref{fig:2}. In other
words, the atoms can be trapped in the area where the magnetic field is
concentrated. For $\kappa =7/3$ the effective trapping potential confines the
atoms tighter compared to the case where $\kappa =1$, as one can see comparing
Figs.~\ref{fig:2}a and \ref{fig:2}b.

Since the effective magnetic field is nearly constant only in a region where $
|\zeta |=\alpha\rho\ll 1$, the effective magnetic flux over this region is much
smaller than its maximum of $ 2\pi\hbar l$, as one can see from
Eq.~(\ref{flux}). In the next subsection we shall show how to produce a strictly
constant magnetic field in the case where $ |\zeta |$ is not necessarily small.

\subsection{Constant effective magnetic field }

\begin{figure}
\begin{center}                                                                  
\includegraphics[width=8.5cm,angle=0]{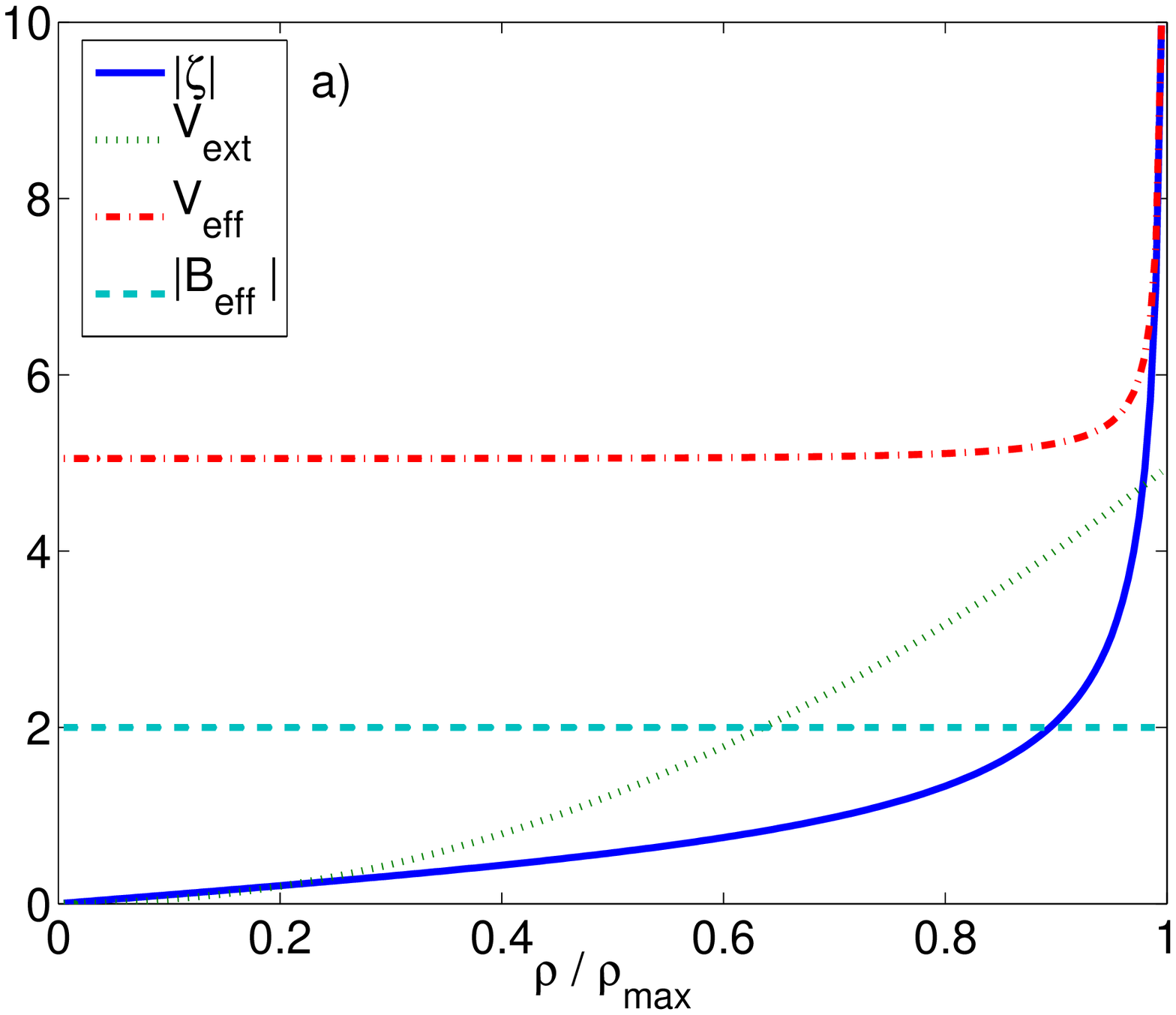}                          
\includegraphics[width=8.5cm,angle=0]{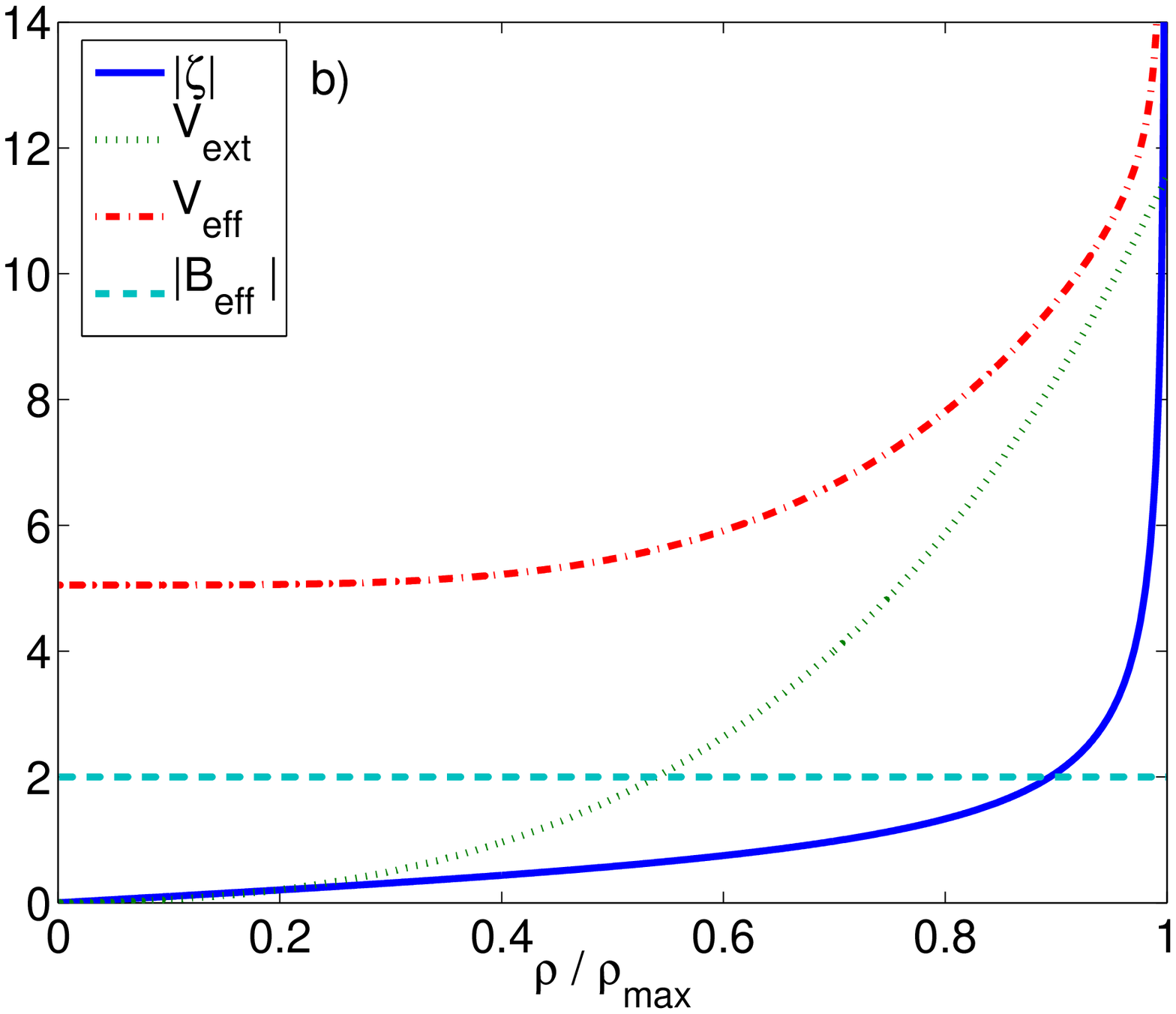}                      
\end{center}
\caption{(Color online) The effective trapping potential $
V_{\mathrm{eff}}$ and the ratio $ |\zeta
|=\left|\Omega_p/\Omega_c\right|$ corresponding to the case where the
effective magnetic field $ B_{\mathrm{eff}}$ is constant. The
external trapping potential $ V_1(\mathbf{r})$ is given by
Eq.~(\ref{V-1-constant}) to compensate the quadratic term in
Eq.~(\ref{V-D-exploding}). The trapping potential for the atoms in the second
hyperfine ground state is chosen to be $ V_2(\mathbf{r})=\kappa
V_1(\mathbf{r})$ with (a) $\kappa =1$ and (b)
$\kappa =7/3$. The other constants are in both cases $
m=\hbar =1$, $ l=10$ and
$\rho_{\mathrm{max}}^2=10$.}
\label{fig:3}
\end{figure}

If we choose 
\begin{equation}
\label{zeta-exploding}
|\zeta |^2=\frac{\left(\rho /\rho_{\mathrm{max}}\right)^2}{1-\left(\rho /\rho_{
\mathrm{max}}\right)^2}\, ,
\end{equation}
the effective vector potential is 
\begin{equation}
\mathbf{A}_{\mathrm{eff}}^{(D)}=-\hbar l\rho\rho_{\mathrm{max}}^{-2}\hat{
\mathbf{e}}_{\phi}\, .
\end{equation}
Consequently we arrive at a constant effective magnetic field 
\begin{equation}
\label{B-constant}
\mathbf{B}_{\mathrm{eff}}=-2\hbar l\rho_{\mathrm{max}}^{-2}
\hat{\mathbf{e}}_z\, ,
\end{equation}
with the corresponding cyclotron frequency
$\omega_c=\hbar 2l/m\rho_{\mathrm{max}}^2$, and the magnetic length
$\ell_B=\sqrt{\hbar /m\omega_c}=\rho_{\mathrm{max}}/\sqrt{2l}$. The effective
trapping potential is now given by 
\begin{equation}
\label{V-D-exploding}
V_{\mathrm{eff}}^{(D)}(\mathbf{r})=V_{\mathrm{ext}}(\mathbf{r})+\frac{\hbar^2}{
2m}\frac{1}{\rho_{\mathrm{max}}^2}\left(l^2d+1/d\right)\, ,
\end{equation}
 where $ d=1-\left(\rho /\rho_{\mathrm{max}}\right)^2$. For
$\rho\rightarrow\rho_{\mathrm{max}}$, the intensity ratio $ |\zeta |^2$ goes to
infinity, so the equations (\ref{zeta-exploding})-(\ref{V-D-exploding}) have a
meaning only for distances smaller than $\rho_{\mathrm{max}}$. Therefore,
Eq.~(\ref{zeta-exploding}) can model an actual intensity distribution of the
control and probe beams only up to a certain radius $\rho_0$ which is smaller
than $\rho_{\mathrm{max}}$. When the radius $\rho_0$ is close to
$\rho_{\mathrm{max}}$, the effective magnetic flux approaches its maximum value
of $ 2\pi\hbar l$.

If 
\begin{equation}
\label{V-1-constant}
V_1(\mathbf{r})=\frac{\hbar^2}{2m\rho_{\mathrm{max}}^2}\left(l^2
-1\right)\left(\rho /\rho_{\mathrm{max}}\right)^2\, ,
\end{equation}
 and $ V_2(\mathbf{r})=\kappa V_1(\mathbf{r})$, the external potential $
V_{\mathrm{ext}}(\mathbf{r})$ given by Eq.~(\ref{vext}) compensates the
quadratic term in Eq.~(\ref{V-D-exploding}). Assuming $\kappa =1$, the overall
effective trapping potential $ V_{\mathrm{eff}}^{(D)}(\mathbf{r})$ is flat
almost up to the large limiting radius $\rho =\rho_{\mathrm{max}}$, as one can
see from Fig.~\ref{fig:3}a. Figure \ref{fig:3}b shows the situation where
$\kappa =7/3$, so that the atoms in hyperfine state $ 2$ are trapped stronger.
In this case, the effective trapping potential becomes tighter. Consequently,
the difference in trapping potentials for different hyperfine states can provide
a natural container confining the trapped atoms within an area of a constant
effective magnetic field.

If the winding number of the probe beam $ l_p=l$ is of the order of a hundred,
the magnetic length $\ell_B=\rho_{\mathrm{max}}/\sqrt{l}$ can be considerably
smaller than the width of an atomic cloud. On the other hand, the diameter of a
pancake shaped cloud is normaly in the range of several tens of micrometers, and
the ratio $\hbar /m$ is of the order of $ 1\mu\mathrm{m}^2/\mathrm{ms}$ for
alkali atoms. Therefore, the cyclotron frequency $\omega_c=\hbar
l/m\rho_{\mathrm{max}}^2$ can be up to several hundreds of Hz which is
comparable to typical trapping frequencies.

\section{\label{sec:concl}Conclusions}

We have considered the influence of two beams of light with orbital angular
momenta on a degenerate gas of electrically neutral atoms (fermions or bosons).
The theory is based on the EIT. We have derived an equation of motion for atoms
driven to a dark state. The equation contains a vector potential type
interaction as well as an effective trapping potential. We have analyzed the
effective vector and trapping potentials in the case where at least one of the
light beams contains an orbital angular momentum. We have shown how to generate
a constant effective magnetic field, as well as a field exhibiting a radial
distance dependence. We have demonstrated that the effective magnetic field can
be concentrated in the area where the effective trapping potential holds the
atoms. In the case of a homogeneous effective magnetic field it is important to
realize that the corresponding cyclotron frequencies and magnetic lengths can be
similar to typical trap frequencies and oscillator lengths used when trapping
cold atoms in BEC and degenerate fermion gases. This will require a high OAM for
the light which is also readily available with present technology.

The theory is based on the adiabatic approximation according to which the atoms
should remain in the dark state. We have estimated that the adiabatic
approximation should hold for atomic velocities up to tens of meters per second,
i.e., up to extremely large velocities in the context of ultra-cold atomic
gases. Such an estimate is lowered if the spontaneous decay of the excited atoms
is taken into account. The atomic dark state accquires then a velocity-dependent
lifetime. For instance, if the atomic velocities are of the order of a
centimeter per second, the atoms should survive in their dark states up to a few
seconds, which is comparable to a typical lifetime of an atomic BEC.

Our proposed method of creating the effective magnetic field has several
advantages compared to a rotating system where only a constant magnetic field is
created \cite{bretin04,schweikhard04,baranov04}. In our method the magnetic
field is shaped and controlled by choosing the proper control and probe beams.
Furthermore stirring an ultra-cold cloud of atoms in a controlled manner is a
rather demanding task, whereas an optically induced vector potential is expected
to be highly controllable.

The theory has already been applied analyzing the de Haas-van Alphen effect in a
gas of electrically neutral atoms \cite{prl04}. It can also be applied to other
intriguing phenomena which intrinsically depend on the magnetic field. For
instance, the quantum Hall effect can now be studied using a cold gas of
electrically neutral atomic fermions. In addition, if the collisional
interaction between the atoms is taken into account, we can study the magnetic
properties of a superfluid atomic Fermi gas \cite{regal04}. Recent advances in
spatial light modulator technology enables us to consider rather exotic light
beams \cite{mcgloin03}. This will allow us to study the effect of different
forms of vector potentials in quantum gases. Finally the combined dynamical
system of light and matter \cite{me02_1} could give an important insight into
gauge theories in general.

\begin{acknowledgments}
This work was supported by the Royal Society of
Edinburgh, the Royal Society of London, the Lithuanian State Science and Studies
Foundation, the Alexander von Humboldt foundation and the Marie-Curie
Trainings-site at the University of Kaiserslautern. Helpful discussions with E.
Andersson, M. Babiker, S. Barnett, J. Courtial, M. Fleisch\-hauer, A.
Kamchatnov, U. Leonhardt, M. Lewenstein, M. Ma\v{s}alas, L. Santos and R.
Unanyan are gratefully acknowledged.
\end{acknowledgments}

\begin{appendix}

\section{\label{sec:app}Equations of motion for $\Phi_D$ and $\Phi_B$}

For derivation of the equations of motion for the dark and bright states
$\Phi_D$ and $\Phi_B$ it is convenient to introduce the notation
\begin{equation}
\xi_c=\frac{1}{\sqrt{1+|\zeta |^2}}\, ,\quad\xi_p=\frac{\zeta}{\sqrt{1+|\zeta
 |^2}}\, .
\end{equation}
To obtain the equation for $\Phi_D$ and $\Phi_B$, let us take the time
derivative of Eq.~(\ref{Psi-D}) and (\ref{Psi-B}) and make use of the original
equations of motion~(\ref{eq-at-g2})-(\ref{eq-at-q2}):
\begin{eqnarray}
i\hbar\dot{\Phi}_D & = & -\frac{\hbar^2}{2m}\left(\xi_c^*\nabla^2\Phi_1
-\xi_p^*\nabla^2\Phi_2\right) +\xi_c^*V_1(\mathbf{r})\Phi_1
 -\xi_p^*\left(\epsilon_{21}+V_2(\mathbf{r})\right)\Phi_2 \nonumber \\
& & +\dot{\xi}_c^*\Phi_1-\dot{\xi}_p^*\Phi_2 \, ,
\label{eq:PsiD-App-A}\\
i\hbar\dot{\Phi}_B & = & -\frac{\hbar^2}{2m}\left(\xi_p\nabla^2\Phi_1
+\xi_c\nabla^2\Phi_2\right) +\xi_pV_1(\mathbf{r})\Phi_1
+\xi_c\left(\epsilon_{21}+V_2(\mathbf{r})\right)\Phi_2 \nonumber\\
& & +\hbar\Omega\frac{\Omega_c^*}{|\Omega_c|}\Phi_3 
+ \dot{\xi}_p\Phi_1+\dot{\xi}_c\Phi_2 \, .
\label{eq:PsiB-App-A}
\end{eqnarray}
Using the inverse transformation 
\begin{equation}
\label{eq:APsi1}
\Phi_1=\frac{1}{\sqrt{1+|\zeta |^2}}(\zeta^*\Phi_B+\Phi_D)
\end{equation}
and 
\begin{equation}
\label{eq:APsi2}
\Phi_2=\frac{1}{\sqrt{1+|\zeta |^2}}(\Phi_B-\zeta\Phi_D)\, ,
\end{equation}
the equations of motion can be represented as 
\begin{equation}
i\hbar\dot{\Phi}_D=\frac{1}{2m}\left(-i\hbar\nabla -\mathbf{A}_{\mathrm{eff}}^{
(D)}\right)^2\Phi_D+V_{\mathrm{eff}}^{(D)}(\mathbf{r})\Phi_D+F_{DB}(\mathbf{
r})\Phi_B\, ,
\end{equation}
and 
\begin{equation}
i\hbar\dot{\Phi}_B=\frac{1}{2m}\left(-i\hbar\nabla +\mathbf{A}_{\mathrm{eff}}^{
(D)}\right)^2\Phi_B+V_{\mathrm{eff}}^{(B)}(\mathbf{r})\Phi_B
+\hbar\Omega\frac{\Omega_c^*}{|\Omega_c|}\Phi_3
+F_{BD}(\mathbf{r})\Phi_D\, ,
\end{equation}
where the effective vector and trapping potentials are explicitly defined by
Eqs.~(\ref{A-eff-D})--(\ref{V-eff-B}) of the main text. The operators $
F_{DB}(\mathbf{r})$ and $ F_{BD}(\mathbf{r})$ describe the transitions between
the dark and bright states:
\begin{eqnarray}
F_{DB}(\mathbf{r})\Phi_B & = &\left[\left(V_1(\mathbf{r})-V_2(\mathbf{r})
-\epsilon_{21}\right)\xi_c^*\xi_p^*+\frac{\hbar^2}{
2m}\left(\xi_p^*\nabla^2\xi_c^*-\xi_c^*\nabla^2\xi_p^*\right)+i\hbar\left(\dot{
\xi}_c^*\xi_p^*-\dot{\xi}_p^*\xi_c^*\right)\right]\Phi_B\nonumber\\
 &  & +\frac{\hbar^2}{m}\left(\xi_p^*\nabla\xi_c^*
-\xi_c^*\nabla\xi_p^*\right)\cdot\nabla\Phi_B\, ,
\label{F-DB}\\
F_{BD}(\mathbf{r})\Phi_D & = &\left[\left(V_1(\mathbf{r})-V_2(\mathbf{r})
-\epsilon_{21}\right)\xi_c\xi_p+\frac{\hbar^2}{2m}\left(\xi_c\nabla^2\xi_p
-\xi_p\nabla^2\xi_c\right)+i\hbar\left(\dot{\xi}_p\xi_c-\dot{
\xi}_c\xi_p\right)\right]\Phi_D\nonumber\\
 &  & +\frac{\hbar^2}{m}\left(\xi_c\nabla\xi_p
-\xi_p\nabla\xi_c\right)\cdot\nabla\Phi_D\, .
\label{F-BD}
\end{eqnarray}
Finally, substituting Eqs.~(\ref{eq:APsi1}) and (\ref{eq:APsi2}) into
Eq.~(\ref{eq-at-e2}), one arrives at Eq.~(\ref{eq:Psi3}) for $\Phi_3$.

\end{appendix}


\begin{thebibliography}{31}
\expandafter\ifx\csname natexlab\endcsname\relax\def\natexlab#1{#1}\fi
\expandafter\ifx\csname bibnamefont\endcsname\relax
  \def\bibnamefont#1{#1}\fi
\expandafter\ifx\csname bibfnamefont\endcsname\relax
  \def\bibfnamefont#1{#1}\fi
\expandafter\ifx\csname citenamefont\endcsname\relax
  \def\citenamefont#1{#1}\fi
\expandafter\ifx\csname url\endcsname\relax
  \def\url#1{\texttt{#1}}\fi
\expandafter\ifx\csname urlprefix\endcsname\relax\def\urlprefix{URL }\fi
\providecommand{\bibinfo}[2]{#2}
\providecommand{\eprint}[2][]{\url{#2}}

\bibitem[{\citenamefont{Davis et~al.}(1995)\citenamefont{Davis, Mewes, Andrews,
  van Druten, Durfee, Kurn, and Ketterle}}]{ketterle95}
\bibinfo{author}{\bibfnamefont{K.~B.} \bibnamefont{Davis}},
  \bibinfo{author}{\bibfnamefont{M.-O.} \bibnamefont{Mewes}},
  \bibinfo{author}{\bibfnamefont{M.~R.} \bibnamefont{Andrews}},
  \bibinfo{author}{\bibfnamefont{N.~J.} \bibnamefont{van Druten}},
  \bibinfo{author}{\bibfnamefont{D.~S.} \bibnamefont{Durfee}},
  \bibinfo{author}{\bibfnamefont{D.~M.} \bibnamefont{Kurn}}, \bibnamefont{and}
  \bibinfo{author}{\bibfnamefont{W.}~\bibnamefont{Ketterle}},
  \bibinfo{journal}{Phys. Rev. Lett.} \textbf{\bibinfo{volume}{75}},
  \bibinfo{pages}{3969} (\bibinfo{year}{1995}).

\bibitem[{\citenamefont{Bradley et~al.}(1995)\citenamefont{Bradley, Sackett,
  Tollett, and Hulet}}]{hulet95}
\bibinfo{author}{\bibfnamefont{C.~C.} \bibnamefont{Bradley}},
  \bibinfo{author}{\bibfnamefont{C.~A.} \bibnamefont{Sackett}},
  \bibinfo{author}{\bibfnamefont{J.~J.} \bibnamefont{Tollett}},
  \bibnamefont{and} \bibinfo{author}{\bibfnamefont{R.~G.} \bibnamefont{Hulet}},
  \bibinfo{journal}{Phys. Rev. Lett.} \textbf{\bibinfo{volume}{75}},
  \bibinfo{pages}{1687} (\bibinfo{year}{1995}).

\bibitem[{\citenamefont{Dalfovo et~al.}(1999)\citenamefont{Dalfovo, Giorgini,
  Pitaevskii, and Stringari}}]{dalfovo99}
\bibinfo{author}{\bibfnamefont{F.}~\bibnamefont{Dalfovo}},
  \bibinfo{author}{\bibfnamefont{S.}~\bibnamefont{Giorgini}},
  \bibinfo{author}{\bibfnamefont{L.}~\bibnamefont{Pitaevskii}},
  \bibnamefont{and}
  \bibinfo{author}{\bibfnamefont{S.}~\bibnamefont{Stringari}},
  \bibinfo{journal}{Rev. Mod. Phys.} \textbf{\bibinfo{volume}{71}},
  \bibinfo{pages}{463} (\bibinfo{year}{1999}).

\bibitem[{\citenamefont{Pitaevskii and Stringari}(2003)}]{bec_stri}
\bibinfo{author}{\bibfnamefont{L.}~\bibnamefont{Pitaevskii}} \bibnamefont{and}
  \bibinfo{author}{\bibfnamefont{S.}~\bibnamefont{Stringari}},
  \emph{\bibinfo{title}{Bose-Einstein Condensation}}
  (\bibinfo{publisher}{Clarendon Press}, \bibinfo{address}{Oxford},
  \bibinfo{year}{2003}).

\bibitem[{\citenamefont{DeMarco and Jin}(1999)}]{demarco99}
\bibinfo{author}{\bibfnamefont{B.}~\bibnamefont{DeMarco}} \bibnamefont{and}
  \bibinfo{author}{\bibfnamefont{D.}~\bibnamefont{Jin}},
  \bibinfo{journal}{Science} \textbf{\bibinfo{volume}{285}},
  \bibinfo{pages}{1703} (\bibinfo{year}{1999}).

\bibitem[{\citenamefont{Schreck et~al.}(2001)\citenamefont{Schreck, Khaykovich,
  Corwin, Ferrari, Bourdel, Cubizolles, and Salomon}}]{salomon01}
\bibinfo{author}{\bibfnamefont{F.}~\bibnamefont{Schreck}},
  \bibinfo{author}{\bibfnamefont{L.}~\bibnamefont{Khaykovich}},
  \bibinfo{author}{\bibfnamefont{K.~L.} \bibnamefont{Corwin}},
  \bibinfo{author}{\bibfnamefont{G.}~\bibnamefont{Ferrari}},
  \bibinfo{author}{\bibfnamefont{T.}~\bibnamefont{Bourdel}},
  \bibinfo{author}{\bibfnamefont{J.}~\bibnamefont{Cubizolles}},
  \bibnamefont{and} \bibinfo{author}{\bibfnamefont{C.}~\bibnamefont{Salomon}},
  \bibinfo{journal}{Phys. Rev. Lett.} \textbf{\bibinfo{volume}{87}},
  \bibinfo{pages}{080403} (\bibinfo{year}{2001}).

\bibitem[{\citenamefont{Hadzibabic et~al.}(2003)\citenamefont{Hadzibabic,
  Gupta, Stan, Schunck, Zwierlein, Dieckmann, and Ketterle}}]{ketterle03}
\bibinfo{author}{\bibfnamefont{Z.}~\bibnamefont{Hadzibabic}},
  \bibinfo{author}{\bibfnamefont{S.}~\bibnamefont{Gupta}},
  \bibinfo{author}{\bibfnamefont{C.~A.} \bibnamefont{Stan}},
  \bibinfo{author}{\bibfnamefont{C.~H.} \bibnamefont{Schunck}},
  \bibinfo{author}{\bibfnamefont{M.~W.} \bibnamefont{Zwierlein}},
  \bibinfo{author}{\bibfnamefont{K.}~\bibnamefont{Dieckmann}},
  \bibnamefont{and} \bibinfo{author}{\bibfnamefont{W.}~\bibnamefont{Ketterle}},
  \bibinfo{journal}{Phys. Rev. Lett} \textbf{\bibinfo{volume}{91}},
  \bibinfo{pages}{160401} (\bibinfo{year}{2003}).

\bibitem[{\citenamefont{Jaksch et~al.}(1998)\citenamefont{Jaksch, Bruder,
  Cirac, Gardiner, and Zoller}}]{dieter98}
\bibinfo{author}{\bibfnamefont{D.}~\bibnamefont{Jaksch}},
  \bibinfo{author}{\bibfnamefont{C.}~\bibnamefont{Bruder}},
  \bibinfo{author}{\bibfnamefont{J.~I.} \bibnamefont{Cirac}},
  \bibinfo{author}{\bibfnamefont{C.~W.} \bibnamefont{Gardiner}},
  \bibnamefont{and} \bibinfo{author}{\bibfnamefont{P.}~\bibnamefont{Zoller}},
  \bibinfo{journal}{Phys. Rev. Lett.} \textbf{\bibinfo{volume}{81}},
  \bibinfo{pages}{3108} (\bibinfo{year}{1998}).

\bibitem[{\citenamefont{Bretin et~al.}(2004)\citenamefont{Bretin, Stock,
  Seurin, and Dalibard}}]{bretin04}
\bibinfo{author}{\bibfnamefont{V.}~\bibnamefont{Bretin}},
  \bibinfo{author}{\bibfnamefont{S.}~\bibnamefont{Stock}},
  \bibinfo{author}{\bibfnamefont{Y.}~\bibnamefont{Seurin}}, \bibnamefont{and}
  \bibinfo{author}{\bibfnamefont{J.}~\bibnamefont{Dalibard}},
  \bibinfo{journal}{Phys. Rev. Lett} \textbf{\bibinfo{volume}{92}},
  \bibinfo{pages}{050403} (\bibinfo{year}{2004}).

\bibitem[{\citenamefont{Schweikhard et~al.}(2004)\citenamefont{Schweikhard,
  Coddington, Engels, Mogendorff, and Cornell}}]{schweikhard04}
\bibinfo{author}{\bibfnamefont{V.}~\bibnamefont{Schweikhard}},
  \bibinfo{author}{\bibfnamefont{I.}~\bibnamefont{Coddington}},
  \bibinfo{author}{\bibfnamefont{P.}~\bibnamefont{Engels}},
  \bibinfo{author}{\bibfnamefont{V.~P.} \bibnamefont{Mogendorff}},
  \bibnamefont{and} \bibinfo{author}{\bibfnamefont{E.~A.}
  \bibnamefont{Cornell}}, \bibinfo{journal}{Phys. Rev. Lett}
  \textbf{\bibinfo{volume}{92}}, \bibinfo{pages}{040404}
  (\bibinfo{year}{2004}).

\bibitem[{\citenamefont{Baranov et~al.}(2005)\citenamefont{Baranov, Osterloh,
  and Lewenstein}}]{baranov04}
\bibinfo{author}{\bibfnamefont{M.~A.} \bibnamefont{Baranov}},
  \bibinfo{author}{\bibfnamefont{K.}~\bibnamefont{Osterloh}}, \bibnamefont{and}
  \bibinfo{author}{\bibfnamefont{M.}~\bibnamefont{Lewenstein}},
  \bibinfo{journal}{Phys. Rev. Lett} \textbf{\bibinfo{volume}{94}},
  \bibinfo{pages}{070404} (\bibinfo{year}{2005}).

\bibitem[{\citenamefont{Jaksch and Zoller}(2003)}]{jaksch03}
\bibinfo{author}{\bibfnamefont{D.}~\bibnamefont{Jaksch}} \bibnamefont{and}
  \bibinfo{author}{\bibfnamefont{P.}~\bibnamefont{Zoller}},
  \bibinfo{journal}{New J. Phys.} \textbf{\bibinfo{volume}{5}},
  \bibinfo{pages}{56} (\bibinfo{year}{2003}).

\bibitem[{\citenamefont{Mueller}(2004)}]{mueller04}
\bibinfo{author}{\bibfnamefont{E.~J.} \bibnamefont{Mueller}},
  \bibinfo{journal}{Phys. Rev. A} \textbf{\bibinfo{volume}{70}},
  \bibinfo{pages}{041603(R)} (\bibinfo{year}{2004}).

\bibitem[{\citenamefont{S{\o}rensen et~al.}(2005)\citenamefont{S{\o}rensen,
  Demler, and Lukin}}]{sorensen04}
\bibinfo{author}{\bibfnamefont{A.~S.} \bibnamefont{S{\o}rensen}},
  \bibinfo{author}{\bibfnamefont{E.}~\bibnamefont{Demler}}, \bibnamefont{and}
  \bibinfo{author}{\bibfnamefont{M.~D.} \bibnamefont{Lukin}},
  \bibinfo{journal}{Phys. Rev. Lett} \textbf{\bibinfo{volume}{94}},
  \bibinfo{pages}{086803} (\bibinfo{year}{2005}).

\bibitem[{\citenamefont{Juzeli{\=u}nas and {\"O}hberg}(2004)}]{prl04}
\bibinfo{author}{\bibfnamefont{G.}~\bibnamefont{Juzeli{\=u}nas}}
  \bibnamefont{and}
  \bibinfo{author}{\bibfnamefont{P.}~\bibnamefont{{\"O}hberg}},
  \bibinfo{journal}{Phys. Rev. Lett} \textbf{\bibinfo{volume}{93}},
  \bibinfo{pages}{033602} (\bibinfo{year}{2004}).

\bibitem[{\citenamefont{Jackiw}(1988)}]{jack03}
\bibinfo{author}{\bibfnamefont{R.}~\bibnamefont{Jackiw}},
  \bibinfo{journal}{Comments At. Mol. Phys.} \textbf{\bibinfo{volume}{21}},
  \bibinfo{pages}{71} (\bibinfo{year}{1988}).

\bibitem[{\citenamefont{Sun and Ge}(1990)}]{Sun90}
\bibinfo{author}{\bibfnamefont{C.-P.} \bibnamefont{Sun}} \bibnamefont{and}
  \bibinfo{author}{\bibfnamefont{M.-L.} \bibnamefont{Ge}},
  \bibinfo{journal}{Phys. Rev. D} \textbf{\bibinfo{volume}{41}},
  \bibinfo{pages}{1349} (\bibinfo{year}{1990}).

\bibitem[{\citenamefont{Dum and Olshanii}(1996)}]{Dum96}
\bibinfo{author}{\bibfnamefont{R.}~\bibnamefont{Dum}} \bibnamefont{and}
  \bibinfo{author}{\bibfnamefont{M.}~\bibnamefont{Olshanii}},
  \bibinfo{journal}{Phys. Rev. Lett.} \textbf{\bibinfo{volume}{76}},
  \bibinfo{pages}{1788} (\bibinfo{year}{1996}).

\bibitem[{\citenamefont{Allen et~al.}(1999)\citenamefont{Allen, Padgett, and
  Babiker}}]{allen99}
\bibinfo{author}{\bibfnamefont{L.}~\bibnamefont{Allen}},
  \bibinfo{author}{\bibfnamefont{M.}~\bibnamefont{Padgett}}, \bibnamefont{and}
  \bibinfo{author}{\bibfnamefont{M.}~\bibnamefont{Babiker}},
  \bibinfo{journal}{Prog. Opt.} \textbf{\bibinfo{volume}{39}},
  \bibinfo{pages}{291} (\bibinfo{year}{1999}).

\bibitem[{\citenamefont{Allen et~al.}(2003)\citenamefont{Allen, Barnett, and
  Padgett}}]{oam}
\bibinfo{author}{\bibfnamefont{L.}~\bibnamefont{Allen}},
  \bibinfo{author}{\bibfnamefont{S.~M.} \bibnamefont{Barnett}},
  \bibnamefont{and} \bibinfo{author}{\bibfnamefont{M.~J.}
  \bibnamefont{Padgett}}, \emph{\bibinfo{title}{Optical Angular Momentum}}
  (\bibinfo{publisher}{Institute of Physics}, \bibinfo{address}{Bristol},
  \bibinfo{year}{2003}).

\bibitem[{\citenamefont{Butts and Rokhsar}(1997)}]{butts97}
\bibinfo{author}{\bibfnamefont{D.~A.} \bibnamefont{Butts}} \bibnamefont{and}
  \bibinfo{author}{\bibfnamefont{D.~S.} \bibnamefont{Rokhsar}},
  \bibinfo{journal}{Phys. Rev. A} \textbf{\bibinfo{volume}{55}},
  \bibinfo{pages}{4346} (\bibinfo{year}{1997}).

\bibitem[{\citenamefont{Mewes et~al.}(2000)\citenamefont{Mewes, Ferrari,
  Schreck, Sinatra, and Salomon}}]{mewes00}
\bibinfo{author}{\bibfnamefont{M.-O.} \bibnamefont{Mewes}},
  \bibinfo{author}{\bibfnamefont{G.}~\bibnamefont{Ferrari}},
  \bibinfo{author}{\bibfnamefont{F.}~\bibnamefont{Schreck}},
  \bibinfo{author}{\bibfnamefont{A.}~\bibnamefont{Sinatra}}, \bibnamefont{and}
  \bibinfo{author}{\bibfnamefont{C.}~\bibnamefont{Salomon}},
  \bibinfo{journal}{Phys. Rev. A} \textbf{\bibinfo{volume}{61}},
  \bibinfo{pages}{011403(R)} (\bibinfo{year}{2000}).

\bibitem[{\citenamefont{Juzeli{\=u}nas and Ma{\v s}alas}(2001)}]{juzeliunas01}
\bibinfo{author}{\bibfnamefont{G.}~\bibnamefont{Juzeli{\=u}nas}}
  \bibnamefont{and} \bibinfo{author}{\bibfnamefont{M.}~\bibnamefont{Ma{\v
  s}alas}}, \bibinfo{journal}{Phys. Rev. A} \textbf{\bibinfo{volume}{63}},
  \bibinfo{pages}{061602(R)} (\bibinfo{year}{2001}).

\bibitem[{\citenamefont{Arimondo}(1996)}]{Arimondo96}
\bibinfo{author}{\bibfnamefont{E.}~\bibnamefont{Arimondo}},
  \bibinfo{journal}{Progr. Opt.} \textbf{\bibinfo{volume}{35}},
  \bibinfo{pages}{259} (\bibinfo{year}{1996}).

\bibitem[{\citenamefont{Harris}(1997)}]{Harris97}
\bibinfo{author}{\bibfnamefont{S.~E.} \bibnamefont{Harris}},
  \bibinfo{journal}{Phys. Today} \textbf{\bibinfo{volume}{50 (7)}},
  \bibinfo{pages}{36} (\bibinfo{year}{1997}).

\bibitem[{\citenamefont{Matsko et~al.}(2001)\citenamefont{Matsko,
  Kocharovskaja, Rostovtsev, Welch, Zibrov, and Scully}}]{eit}
\bibinfo{author}{\bibfnamefont{A.~B.} \bibnamefont{Matsko}},
  \bibinfo{author}{\bibfnamefont{O.}~\bibnamefont{Kocharovskaja}},
  \bibinfo{author}{\bibfnamefont{Y.}~\bibnamefont{Rostovtsev}},
  \bibinfo{author}{\bibfnamefont{G.~R.} \bibnamefont{Welch}},
  \bibinfo{author}{\bibfnamefont{A.~S.} \bibnamefont{Zibrov}},
  \bibnamefont{and} \bibinfo{author}{\bibfnamefont{M.~O.}
  \bibnamefont{Scully}}, \bibinfo{journal}{Advances in Atomic, Molecular, and
  Optical Physics} \textbf{\bibinfo{volume}{46}}, \bibinfo{pages}{191}
  (\bibinfo{year}{2001}).

\bibitem[{\citenamefont{Lukin}(2003)}]{lukin03}
\bibinfo{author}{\bibfnamefont{M.~D.} \bibnamefont{Lukin}},
  \bibinfo{journal}{Rev. Mod. Phys.} \textbf{\bibinfo{volume}{75}},
  \bibinfo{pages}{457} (\bibinfo{year}{2003}).

\bibitem[{\citenamefont{Hau et~al.}(1999)\citenamefont{Hau, Harris, Dutton, and
  Behrooz}}]{hau99}
\bibinfo{author}{\bibfnamefont{L.~V.} \bibnamefont{Hau}},
  \bibinfo{author}{\bibfnamefont{S.~E.} \bibnamefont{Harris}},
  \bibinfo{author}{\bibfnamefont{Z.}~\bibnamefont{Dutton}}, \bibnamefont{and}
  \bibinfo{author}{\bibfnamefont{C.}~\bibnamefont{Behrooz}},
  \bibinfo{journal}{Nature} \textbf{\bibinfo{volume}{397}},
  \bibinfo{pages}{594} (\bibinfo{year}{1999}).

\bibitem[{\citenamefont{Regal et~al.}(2004)\citenamefont{Regal, Greiner, and
  Jin}}]{regal04}
\bibinfo{author}{\bibfnamefont{C.~A.} \bibnamefont{Regal}},
  \bibinfo{author}{\bibfnamefont{M.}~\bibnamefont{Greiner}}, \bibnamefont{and}
  \bibinfo{author}{\bibfnamefont{D.~S.} \bibnamefont{Jin}},
  \bibinfo{journal}{Phys. Rev. Lett.} \textbf{\bibinfo{volume}{92}},
  \bibinfo{pages}{040403} (\bibinfo{year}{2004}).

\bibitem[{\citenamefont{McGloin et~al.}(2003)\citenamefont{McGloin, Spalding,
  Melville, Sibbett, and Dholakia}}]{mcgloin03}
\bibinfo{author}{\bibfnamefont{D.}~\bibnamefont{McGloin}},
  \bibinfo{author}{\bibfnamefont{G.}~\bibnamefont{Spalding}},
  \bibinfo{author}{\bibfnamefont{H.}~\bibnamefont{Melville}},
  \bibinfo{author}{\bibfnamefont{W.}~\bibnamefont{Sibbett}}, \bibnamefont{and}
  \bibinfo{author}{\bibfnamefont{K.}~\bibnamefont{Dholakia}},
  \bibinfo{journal}{Opt. Express} \textbf{\bibinfo{volume}{11}},
  \bibinfo{pages}{158} (\bibinfo{year}{2003}).

\bibitem[{\citenamefont{{\"O}hberg}(2002)}]{me02_1}
\bibinfo{author}{\bibfnamefont{P.}~\bibnamefont{{\"O}hberg}},
  \bibinfo{journal}{Phys. Rev. A} \textbf{\bibinfo{volume}{66}},
  \bibinfo{pages}{021603(R)} (\bibinfo{year}{2002}).

\end{thebibliography}
\end{document}